\documentstyle[epsf,aps]{revtex}
\begin{document}
\begin{flushright}
NUC-MN-99/2-T \\
TPI-MINN-99/10 \\
UMN-TH-99/1746
\end{flushright}
\vspace*{1cm}
\setcounter{footnote}{1}
\begin{center}
  {\Large\bf Diffractive Structure Function in a Quasi--Classical
    Approximation} \\[1cm] Yuri V.\ Kovchegov and Larry McLerran \\ ~~
  \\ {\it School of Physics and Astronomy, University of Minnesota, \\
    Minneapolis, MN 55455 }\\ ~~ \\ ~~ \\
\end{center}
\begin{abstract}
  We derive an expression for diffractive $F_2$ structure function
  which should be valid at small $x$ for quasi--elastic scattering on
  a hadron and for quasi-elastic scattering on a large nucleus.  This
  expression includes multiple rescatterings of the quark--antiquark
  pair produced by the virtual photon off the sources of color charge
  in a quasi--classical approximation.  We find that there is a
  relation between such diffractive production and inclusive
  processes.  In the former, one averages over all colors of sources
  before squaring the amplitude, and in the latter one first squares
  the amplitude and then averages it in the hadron's or nuclear wave
  function. We show that in the limit of a large virtuality of the
  photon $Q^2$ the diffractive structure function becomes linearly
  proportional to the gluon distribution of the hadron or nucleus,
  therefore proving that in this sense diffraction is a leading twist
  effect.
\end{abstract}

\section{Introduction}

\begin{figure}
\begin{center}
\epsfxsize=12cm
\leavevmode
\hbox{ \epsffile{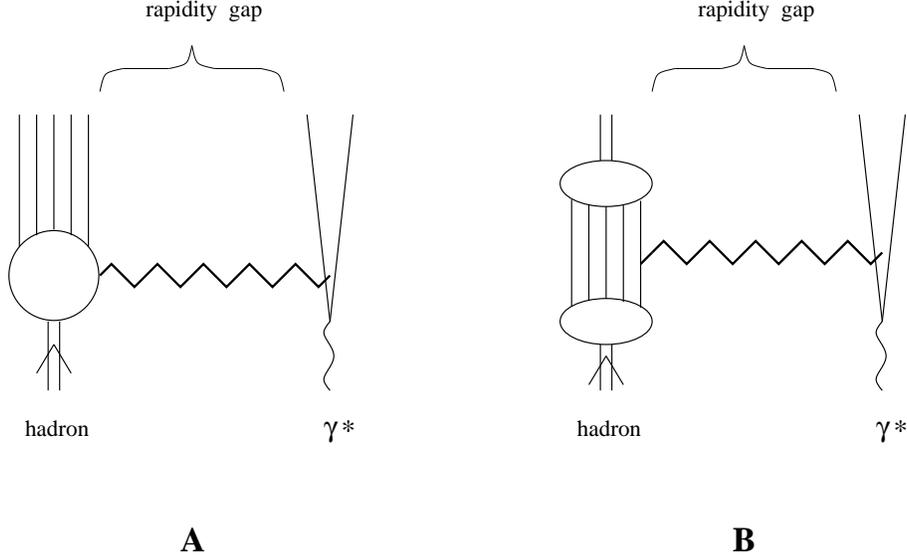}}
\end{center}
\caption{(A) Diffractive deep inelastic scattering with a rapidity gap. 
  (B) The same process with the hadron remaining intact in the final
  state (quasi--elastic photoproduction). The interaction is roughly
  illustrated by an exchange of some color singlet particle
  (pomeron).}
\label{gap}
\end{figure}

In recent papers by Buchm\"{u}ller, Hebecker and collaborators
\cite{Heb,BDH,BGH} diffractive processes are considered where a
virtual photon scatters on a hadron or a nucleus, producing two jets
out of a quark and an antiquark in the virtual photon's wave function,
and a rapidity gap between the quark-antiquark pair and the hadron
(nucleus) with which the virtual photon scatters. In the final state
the hadron (nucleus) forms a cluster of diffractively produced soft
particles. The process is illustrated in Fig. \ref{gap}a.
Buchm\"{u}ller, Hebecker and colleagues \cite{Heb,BDH,BGH} make the
provocative claim that such processes may be computed in a
semi-classical description of small $x$ processes where the effect of
the hadron is taken into account as classical sources of color charge
\cite{MV,MV1,k,KR,kmmw,klmw,JKW}.  They claim that such diffractive
processes may be computed by first computing the amplitude of the
process, then averaging it over the various color orientations of the
source and finally squaring the amplitude to obtain the cross section.
That way the squaring of the amplitude and averaging in the hadron's
(nuclear) wave function are done in the order opposite to what one
usually does when calculating total (inclusive) cross section.

In this same theory \cite{Heb,BDH,BGH}, the structure functions for
deep inelastic scattering are computed using the same distribution of
classical sources.  The only difference is that the amplitude is first
squared and then averaged over color.  If these prescriptions are
true, then there is a subtle and deep relation between the structure
functions computed in deep inelastic scattering and diffractive
virtual photon production of jets.

In this paper, we find that we can derive this relation for a
restricted class of diffractive processes with a rapidity gap where
the hadron encountered remains intact (see Fig. \ref{gap}b).  We shall
call such processes quasi-elastic virtual photoproduction.  Although
it might be possible to extend our methods to the more general case
where the hadron fragments into soft particles, we are unable to do so
with our current technique.

We shall present two types of arguments.  The first is an intuitive
argument similar to that invoked in Glauber scattering.  We shall
pursue the argument in the context of a large nucleus where one has a
simple picture of the sources of color charge arising from the nucleon
valence quark distribution.  We will then later generalize the
argument to a hadron where the sources are largely gluonic and arise
from a proper renormalization group treatment of the hadronic
wavefunction.

We shall perform this analysis when the averaging over color sources
is Gaussian in the same sense as was discussed in
\cite{MV,MV1,k,KR,kmmw,klmw,JKW}, or more generally local in
transverse space.  If the renormalization group analysis after
including the QCD evolution in rapidity \cite{kmmw,klmw,JKW} yields
the weight function for averaging over sources which is not local, as
it well might be, then our prescription would fail to impose the
condition of a diffractive rapidity gap on the process at hand and
should be understood in the sense of quasi--classical approximation
only.

We derive explicit expressions for the quasi--elastic structure
function of a nucleus or a hadron.  This structure function is
analogous of the diffractive structure function of Buchm\"{u}ller,
Hebecker and colleagues.  We shall refer to it as $F_2^D$ with the
caveat that we are capable at present of describing those processes
which are quasi--elastic for the hadron.  We also derive an expression
for the structure function for deep inelastic scattering $F_2$,
showing that $F_2$ and $F_2^D$ are related by an interchange of color
averaging and squaring of the amplitude.

We analyze the large $Q^2$ limit of the expression we obtain for the
diffractive structure function $F_2^D$. The result is very
interesting. The diffractive structure function of the hadron or
nucleus turns out to be linearly proportional to the gluon
distribution of the hadron or nucleus: $F_2^D (x_{Bj}, Q^2) \sim
x_{Bj} G (x_{Bj}, Q^2)$. That way we prove that due to the multiple
rescattering effects diffraction becomes a leading twist effect
\cite{kope}, in the sense that diffractive structure function is not
suppressed by extra powers of $Q^2$ with respect to the total
structure function. At large enough $Q^2$ diffractive structure
function should depend on center of mass energy for deep inelastic
scattering, or, equivalently, on Bjorken $x_{Bj}$, in the same way as
the structure function $F_2$. This statement is confirmed by recent
HERA data \cite{zeus}.

The outline of the paper is the following: In Sect. II we construct
the light cone wavefunction of a virtual photon.  This will be useful
as we shall derive the scattering amplitude by convoluting this
wavefunction with the propagator of the quark--antiquark pair through
the hadron which includes multiple scatterings of the $q {\overline
  q}$ pair on the gluons (classical sources) which constitute the
hadron.

In Sect. III we analyze diffractive scattering from a large nucleus
and derive an expression for $F_2$.  This example illustrates all the
essential points of the more general analysis where a hadron is
treated as a sum of sources arising from a renormalization group
treatment of the hadron wavefunction. We show how in the large $Q^2$
limit due to the multiple rescattering effects diffractive structure
function behaves like a leading twist expression.

In Sect. IV we generalize scattering off a nucleus to a hadron.  We
argue that if the weight function which is used to color average the
sources is Gaussian the result of the previous analysis are
essentially reproduced.  If the averaging procedure is local (but not
Gaussian) we shall see how the result above generalizes.  We have not
been able to generalize to a non-local (in transverse coordinates)
averaging procedure.

In Sect. V we summarize our results, and compare with the case of deep
inelastic scattering by deriving expressions for the total inclusive
cross--section and $F_2$ in the quasi--classical limit.

\section{Light Cone Wave Function of a Virtual Photon}

\begin{figure}
\begin{center}
\epsfxsize=10cm
\leavevmode
\hbox{ \epsffile{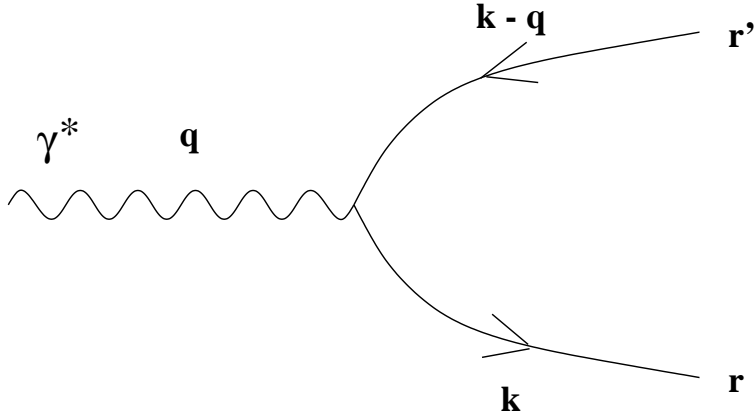}}
\end{center}
\caption{Light cone wave function of a virtual photon.}
\label{wf}
\end{figure}

Here we are going to calculate the wave function of quark--antiquark
fluctuations of a virtual photon. The diagram is shown in Fig.
\ref{wf}. Without any loss of generality we can work in a frame where
the transverse momentum of the virtual photon is zero, so that the
momentum of the photon is given by $q_\mu = \left( q_+ , -
  \frac{Q^2}{2 q_+} , \bf{0} \right)$ with $Q^2$ the virtuality of the
photon. The light cone momentum of the virtual photon $q_+$ is very
large, so that in the spirit of the eikonal approximation we will
neglect all the inverse powers of $q_+$ in the calculations below.
However we will keep all the inverse powers of $Q^2$, therefore
resumming all the ``higher twist'' terms.

Using the light cone perturbation theory with the Feynman rules from
\cite{Bro} one can write down the following expressions for the value
of the amplitude in Fig.  \ref{wf}:
\begin{equation}
  \psi_{r r'}^\lambda ({\bf k}, z) = e_f \, \frac{\sqrt{z (1-z)}}{{\bf
      k}^2 + m_f^2 + Q^2 z (1 - z)} \tilde{u}_r (q - k) \,
  \epsilon^{(\lambda)} \cdot \gamma \, v_{r'} (k)
\end{equation}
with $r$ and $r'$ being quark and antiquark helicities
correspondingly, $r = \pm 1$, $\lambda$ is the photon's polarization,
$\lambda = \pm 1$, $m_f$ is the quark's mass and $e_f$ is its electric
charge ($f$ denotes quark's flavor). We defined $z = k_+ / q_+ $ as
the fraction of the photon's light cone momentum carried by the quark.
We will calculate separately the cases of a transverse and
longitudinal polarization of the photon.  We start with transverse
polarization $\epsilon_{T, \mu}^{\lambda} =
(0,0,{\bf\epsilon}^\lambda)$, with $\epsilon^\lambda = ( 1 + \lambda
i) / \sqrt{2} $. After plugging in the explicit expressions for
$\tilde{u}_r$ and $v_{r'}$ a straightforward calculation yields
\begin{equation}\label{momwf}
  \psi_{r r'}^{T \, \lambda} ({\bf k}, z) = e_f \frac{1}{{\bf k}^2 +
    m_f^2 + Q^2 z (1 - z)} \left( \delta_{r r'} {\bf k} \cdot
    {\bf\epsilon}^\lambda [r (1 - 2 z) + \lambda ] + r \delta_{r, -r'}
    m_f (1 + r \lambda) \right).
\end{equation}
For the reasons which will become apparent later we want to obtain the
virtual photon's wave function in transverse coordinate space.
Performing a Fourier transform of Eq. (\ref{momwf})
\begin{equation}
  \psi_{r r'}^{\lambda} ({\bf x}, z) = \int \frac{d^2 k}{(2
    \pi)^2} e^{- i {\bf k} \cdot {\bf x}} \psi_{r r'}^{\lambda}
  ({\bf k}, z)
\end{equation}
we obtain
\begin{equation}
  \psi_{r r'}^{T \, \lambda} ({\bf x}, z) = \frac{e_f}{2 \pi} \left(
    \delta_{r r'} i {\bf\epsilon}^\lambda \cdot \nabla_x [r (1 - 2 z)
    + \lambda ] + r \delta_{r, -r'} m_f (1 + r \lambda) \right) K_0 (x
  a),
\end{equation}
where $a^2 = Q^2 z (1 - z) + m^2_f$ and $x = |{\bf x}|$. Defining
\begin{equation}
  \Phi_T ({\bf x},{\bf y},z) = \frac{N_c}{2} \sum_{\lambda, r, r', f}
  \psi_{r r'}^{T \, \lambda} ({\bf x}, z) \, \psi_{r r'}^{* \,T \,
    \lambda } ({\bf y}, z)
\end{equation}
we derive
\begin{equation}\label{trsq}
  \Phi_T ({\bf x},{\bf y},z) = 2 N_c \sum_f \frac{\alpha^f_{EM}}{\pi}
  \left( a^2 \frac{{\bf x} \cdot {\bf y}}{x y} K_1 (x a) K_1 (y a)
    [z^2 + (1 - z)^2] + m_f^2 K_0 (x a) K_0 (y a) \right).
\end{equation}
When ${\bf x} ={\bf y}$ the object in Eq. (\ref{trsq}) becomes just a
square of the transverse virtual photon's wave function.

To calculate the longitudinal contribution to the wave function we
write down an expression for the longitudinal polarization by
requiring that $\epsilon^2 = 1$ and $\epsilon \cdot q = 0$. We end up
with $\epsilon_{L,\mu}^\lambda = \left( \frac{q_+}{Q}, \frac{Q}{2
    q_+}, {\bf 0} \right)$. Similarly to the above one obtains
\begin{equation}\label{momlwf}
  \psi^L_{r r'} ({\bf k},z) = e_f \frac{1}{{\bf k}^2 + m_f^2 + Q^2 z
    (1-z)} 2 Q \delta_{r r'} z (1-z).
\end{equation}
Fourier transformation of Eq. (\ref{momlwf}) yields
\begin{equation}
  \psi^L_{r r'} ({\bf x},z) = \frac{e_f}{2 \pi} \delta_{r r'} z (1-z)
  2 Q K_0 (x a),
\end{equation}
which results in
\begin{equation}\label{losq}
  \Phi_L ({\bf x},{\bf y},z) = 2 N_c \sum_f \frac{\alpha^f_{EM}}{\pi}
  \left[ 4 Q^2 z^2 (1 - z)^2 K_0 (x a) K_0 (y a) \right].
\end{equation}
Eqs. (\ref{trsq}) and (\ref{losq}) provide us with the square of the
virtual photon's wave function:
\begin{equation}\label{phieq}
  \Phi ({\bf x},{\bf y},z) = \Phi_T ({\bf x},{\bf y},z) + \Phi_L ({\bf
    x},{\bf y},z).
\end{equation}

\section{Diffractive Structure Function}

In the calculation of the quark--antiquark propagator through the
nucleus we will make use of the quasi--classical approximation
\cite{mM,BDMPS}, which restricts the interactions of the $q
\overline{q}$ pair with the nucleons to nothing more than a two gluon
exchange with the nucleon in covariant gauge.  Light cone ($A_- = 0$)
gauge analysis of the process is a bit more saddle, but leads to the
same formulation of the classical limit \cite{k}. 

\begin{figure}
\begin{center}
\epsfxsize=12cm
\leavevmode
\hbox{ \epsffile{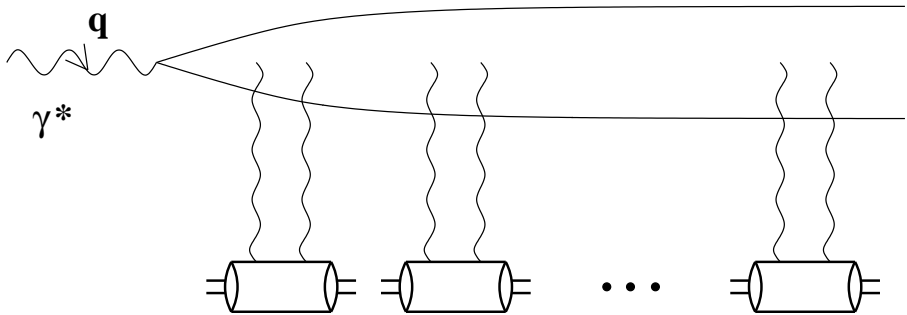}}
\end{center}
\caption{Diffractive quark--antiquark production in DIS.}
\label{ampl}
\end{figure}

\begin{figure}
\begin{center}
\epsfxsize=12cm
\leavevmode
\hbox{ \epsffile{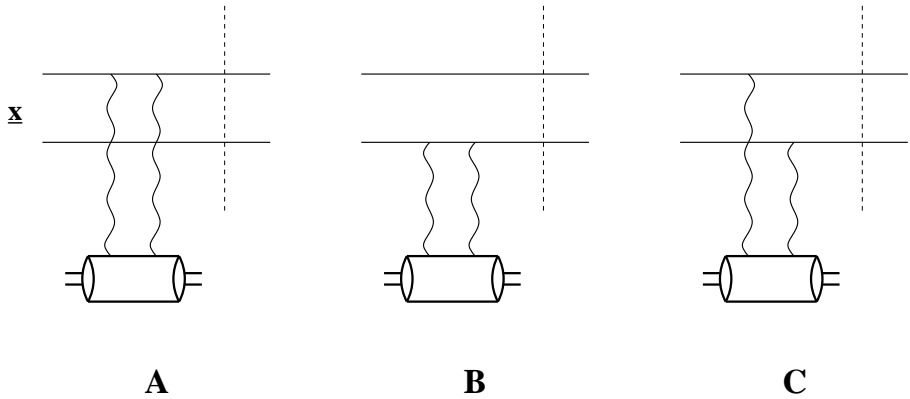}}
\end{center}
\caption{Calculation of the quark--antiquark propagator.}
\label{prop}
\end{figure}

As was stated above, in order to calculate diffractive (or elastic)
production of the quark--antiquark pair one has to average the
production amplitude in the nuclear wave function.  After one does
that all the one gluon exchange contributions would cancel resulting
in two gluon exchange (virtual) interactions being the only
possibility. The amplitude is depicted in Fig. \ref{ampl}.  There the
gluon lines can hook in all possible ways to the quark lines (see Fig.
\ref{prop} ). Similarly to the wave functions, the propagator is
calculated in the eikonal approximation, i.e., the $q \overline{q}$
pair is assumed to have a very large light cone momentum, inverse
powers of which in the amplitude will be neglected.

In order to calculate the quark--antiquark propagator we have to start
at the lowest order case of one nucleon by summing the diagrams shown
in Fig. \ref{prop}. The vertical dashed line there indicates the final
state of the $q \overline{q}$ pair. Because the color averaging is
done in the amplitude and the complex conjugate amplitude
independently, the propagators in the amplitude and the complex
conjugate amplitude effectively factorize. That way we can do the
summation just in the amplitude and the answer for the complex
conjugate amplitude will be the same.  That is why in Fig. \ref{prop}
we show the lowest order interactions only in the amplitude.

To evaluate diagrams in Fig. \ref{prop}, similarly to \cite{mM,BDMPS}
we start by defining the normalized quark--nucleon scattering
cross--section
\begin{equation}
  V ({\bf l}) = \frac{1}{\sigma} \frac{d \sigma}{d^2 {\bf l}}
\end{equation}
where ${\bf l}$ is the momentum transfer. The Fourier transformation of
$V ({\bf l})$ is defined by
\begin{equation}
  \tilde{V} ({\bf x}) = \int d^2 l \, e^{- i {\bf l} \cdot {\bf x}} V
  ({\bf l}).
\end{equation}
Then, as one can easily see, the diagrams in Fig. \ref{prop} give the
following factors: A and B give $- (1/2) \tilde{V} ({\bf 0})$ each,
and C gives $\tilde{V} ({\bf x})$, where ${\bf x} $ is the transverse
separation of the pair. Defining
\begin{equation}
\tilde{v} ({\bf x}) = \frac{4}{x^2} [1 - \tilde{V} ({\bf x})]
\end{equation}
and adding all the contributions of graphs in Fig. \ref{prop} we
obtain
\begin{equation}
  A + B + C = - [\tilde{V} ({\bf 0}) - \tilde{V} ({\bf x})] = -
  \frac{1}{4} {\bf x}^2 \tilde{v} ({\bf x}).
\end{equation}
Now to obtain a contribution to the propagator we have to multiply
this result by $\rho \sigma = 1 / \lambda$, to take into account the
density of nucleons and quark--nucleon cross-section, and by the
length of the path along the $z$ direction of the quark--antiquark
pair in the nuclear matter $L$. $\lambda$ is the mean free path of the
pair in the nucleus. One can easily see that if we start including
more and more nucleons, interacting with the $q \overline{q}$ pair,
the answer for the propagator will be just an exponentiation of the
lowest order result. Since in light cone perturbation theory one has
to also include the diagram without interactions, the final answer for
the propagator of the pair in the nuclear matter will be
\begin{equation}
  \exp \left( - \frac{{\bf x}^2 \tilde{v} ({\bf x}) L}{4 \lambda}
  \right).
\end{equation}
One should note here that our convention is different from \cite{mM},
since $\sigma$ in our case is the quark--nucleon cross--section, same
as in \cite{BDMPS}.

An important fact to note is that in the two gluon exchange, no net
transverse momentum is carried to the source.  The gluon coming in has
the same transverse momentum as the gluon going out.  Additionally,
the longitudinal momentum transfer is very small, of order $e^{-\Delta
  y}$ where $\Delta y$ is the rapidity gap between the source and the
quark-antiquark pair.  This rapidity gap will in general be large.
Therefore the source barely changes its momentum in the interaction.

In the language of the Fock space wavefunctions one can see that the
Fock space wavefunction of the target nucleus is essentially the same
after the interaction as it was before.  Therefore with probability
close to one, the hadronic wavefunction is unchanged.  Therefore the
procedure of averaging over the sources before squaring the amplitude
projects out the quasi-elastic piece of the interaction.  In the final
state there is a quark-antiquark jet, a large rapidity gap, and
finally a nucleus with its momentum largely unchanged.

Of course we expect that a real nucleus having small binding energies
will fragment.  Presumably quasi-elastic means here only no inelastic
particle production in the central rapidity region.  On the other hand
for a real hadron, where we imagine the sources as gluons arising from
a proper renormalization group treatment of the hadron wavefunction,
quasi-elastic means no fragmentation of the hadron.

This leaves open the question of how one treats truly diffractive
processes where one allows for some fragmentation of the hadron into
pions which have momentum close to the hadron's initial momentum.
Presumably one can find a relationship to the quasi-elastic
calculation, but we have not been able to do so. 

In order to write down the final answer for the $q \overline{q}$
production cross--section it is convenient to work in the rest frame
of the nucleus. Analogous to \cite{mM} we assume that the virtual
photon (or the quark--antiquark pair) reaches the nucleus at light
cone time $\tau = x_- =0$. Then, if we define $\tau_1$ as the time
when the virtual photon splits into the quark--antiquark pair in the
amplitude and $\tau_2$ as the time when the virtual photon splits into
the quark--antiquark pair in the complex conjugate amplitude, one
should consider four cases:
\begin{mathletters}
\begin{equation}
  (a) \hspace{1cm} \tau_1 > 0 , \, \tau_2 > 0
\end{equation}
\begin{equation}
  (b) \hspace{1cm} \tau_1 < 0 , \, \tau_2 > 0
\end{equation}
\begin{equation}
  (c) \hspace{1cm} \tau_1 > 0 , \, \tau_2 < 0
\end{equation}
\begin{equation}
  (d) \hspace{1cm} \tau_1 < 0 , \, \tau_2 < 0
\end{equation}
\end{mathletters}
The corresponding cross--sections will be
\begin{mathletters}\label{diffxss}
\begin{equation}
  \frac{d \sigma^{(a)}}{d^2 k \, dz} = \frac{1}{\pi (2 \pi)^3} \, \int
  d^2 b \, d^2 x \, d^2 y \,e^{- i {\bf k} \cdot ({\bf x} - {\bf y})}
  \, \Phi ({\bf x},{\bf y}, z )
\end{equation}
\begin{equation}
  \frac{d \sigma^{(b)}}{d^2 k \, dz} = - \frac{1}{\pi (2 \pi)^3} \,
  \int d^2 b \, d^2 x \, d^2 y \,e^{- i {\bf k} \cdot ({\bf x} - {\bf
      y})} \, \Phi ({\bf x},{\bf y}, z ) \, \exp \left( - \frac{{\bf
        x}^2 \tilde{v} ({\bf x}) \sqrt{R^2 - b^2}}{2 \lambda} \right)
\end{equation}
\begin{equation}
  \frac{d \sigma^{(c)}}{d^2 k \, dz} = - \frac{1}{\pi (2 \pi)^3} \,
  \int d^2 b \, d^2 x \, d^2 y \,e^{- i {\bf k} \cdot ({\bf x} - {\bf
      y})} \, \Phi ({\bf x},{\bf y}, z ) \, \exp \left( - \frac{{\bf
        y}^2 \tilde{v} ({\bf y}) \sqrt{R^2 - b^2}}{ \lambda} \right)
\end{equation}
\begin{equation}
  \frac{d \sigma^{(d)}}{d^2 k \, dz} = \frac{1}{\pi (2 \pi)^3} \, \int
  d^2 b \, d^2 x \, d^2 y \,e^{- i {\bf k} \cdot ({\bf x} - {\bf y})}
  \, \Phi ({\bf x},{\bf y}, z ) \, \exp \left( - \frac{{\bf x}^2
      \tilde{v} ({\bf x}) \sqrt{R^2 - b^2}}{2 \lambda} \right) \, \exp
  \left( - \frac{{\bf y}^2 \tilde{v} ({\bf y}) \sqrt{R^2 - b^2}}{2
      \lambda} \right)
\end{equation}
\end{mathletters}
with $\Phi ({\bf x},{\bf y}, z )$ given by Eq. (\ref{phieq}). Here $R$
is the radius of the nucleus, which we assume to be spherical, and $b$
is the impact parameter. Adding together the contributions given by
Eqs.  (\ref{diffxss}) we find the expression for the quark--antiquark
production cross--section in deep inelastic scattering on a nucleus:
\begin{eqnarray*}
  \frac{d \sigma^D}{d^2 k \, dz} = \frac{1}{\pi (2 \pi)^3} \, \int d^2
  b \, d^2 x \, d^2 y \,e^{- i {\bf k} \cdot ({\bf x} - {\bf y})} \,
  \Phi ({\bf x},{\bf y}, z )
\end{eqnarray*}
\begin{eqnarray}\label{diffxs}
  \times \left[ 1 - \exp \left( - \frac{{\bf x}^2 \tilde{v} ({\bf x})
        \sqrt{R^2 - b^2}}{2 \lambda} \right) \right] \, \left[ 1 -
    \exp \left( - \frac{{\bf y}^2 \tilde{v} ({\bf y}) \sqrt{R^2 -
          b^2}}{2 \lambda} \right) \right].
\end{eqnarray}
Making use of the definition of the $F_2$ structure function
\begin{equation}
F_2 = \frac{Q^2}{4 \pi^2 \alpha_{EM}} (\sigma_T + \sigma_L)
\end{equation}
together with Eqs. (\ref{trsq}),(\ref{losq}),(\ref{phieq}), we derive
the following formula for the diffractive structure function
\begin{eqnarray*}
  F_2^D = Q^2 \frac{N_c}{\pi (2 \pi)^3} \sum_f \, Z^2_f \, \int d^2 b
  \, d^2 x \, \int_0^1 dz \left\{ a^2 K_1^2 (x a) [z^2 + (1 - z)^2] +
    m_f^2 K_0^2 (x a) + 4 Q^2 z^2 (1 - z)^2 K_0^2 (x a) \right\}
\end{eqnarray*}
\begin{eqnarray}\label{diffsf}
  \times \left[ 1 - \exp \left( - \frac{{\bf x}^2 \tilde{v} ({\bf x})
        \sqrt{R^2 - b^2}}{2 \lambda} \right) \right]^2 ,
\end{eqnarray}
where $Z_f^2 = \alpha_{EM}^f / \alpha_{EM}$ for each quark's flavor.
Here we should note that in the logarithmic approximation of small
${\bf x}$ (large $Q^2$)
\begin{equation}\label{vg}
  \frac{\tilde{v} ({\bf x})}{\lambda} = \frac{2 \pi^2 \alpha_s}{N_c}
  \, \rho \, x_{Bj} G (x_{Bj} , 1/{\bf x}^2)
\end{equation}
(see Eq. (4.7) of \cite{BDMPS} as well as \cite{Genya}). Now
$\alpha_s$ is the strong coupling constant, $\rho$ is the density of
nuclear matter, normalized to give the atomic number $A$ when
integrated over the whole nucleus. $x_{Bj} G (x_{Bj} , Q^2)$ is the
gluon distribution in a nucleon, which we take at the lowest order in
$\alpha_s$ (see the next section). Using Eq. (\ref{vg}) one can see
now that in the large $Q^2$ limit the differential cross section of
Eq. (\ref{diffxs}) becomes proportional to the square of the gluon
distribution function of the hadron or nucleus, reducing to an old and
well known result \cite{g2}.

However, the large $Q^2$ limit of $F_2^D$ is more interesting. Let us
assume for simplicity of the calculations (only in this section) that
our target is a large cylindrical nucleus with the radius $R$ and the
length $2 R$ along the axis. Then the impact parameter integration in
Eq.(\ref{diffsf}) becomes trivial giving just a pre-factor of $\pi
R^2$.  Using the representation of the modified Bessel function $K_1
(\xi)$ \footnote{We thank D. Kharzeev for showing us this trick.}
\begin{equation}
  K_1 (\xi) = \frac{\xi}{4} \int_0^\infty \frac{d t}{t^2} \, e^{- t -
    \frac{\xi^2}{4 t}}
\end{equation}
  in Eq. (\ref{diffsf}), assuming that the dominating values of $z$
  are small, performing the $z$ and $t$ integrations, we obtain
\begin{eqnarray}\label{laq1}
  F_2^D = \frac{N_c R^2}{3 \pi^2} \sum_f \, Z^2_f
  \,\int_{1/Q^2}^\infty \frac{d x^2}{(x^2)^2} \left[ 1 - \exp \left( -
      \frac{\alpha_s \pi}{2 N_c R^2} x^2 x_{Bj} G (x_{Bj} , 1/x^2)
    \right) \right]^2 ,
\end{eqnarray}
where $x_{Bj} G (x_{Bj} , Q^2)$ is now the gluon distribution function
of the whole nucleus.  In the leading logarithmic approximation we
assume that the gluon distribution is a slowly varying function of
$Q^2$.  That way it will effectively play the role of the upper cutoff
of the $x$ integration in Eq. (\ref{laq1}). Performing the integration
in Eq.  (\ref{laq1}) we obtain the following expression for the
diffractive structure function at large $Q^2$ :
\begin{equation}\label{laq}
  F_2^D (x_{Bj} , Q^2) = \frac{\alpha_s}{6 \pi}\sum_f \, Z^2_f \,
  x_{Bj} G (x_{Bj} , Q^2).
\end{equation}
A little more careful estimates would introduce a factor of $2 \ln 2$
in front of $x_{Bj} G$ on the right hand side of (\ref{laq}). One
should note here that the exact coefficient can not be precisely fixed
by this method, since to be able to use $x_{Bj} G$ as a cutoff in Eq.
(\ref{laq1}) one should reach the kinematic region of large gluon
distributions. This would correspond to evolving $x_{Bj} G (x_{Bj} ,
Q^2)$ with some QCD evolution. However, in the strict sense, our
$x_{Bj} G$ should be taken only at the lowest order in $\alpha_s$ [see
Eq. (\ref{lo}) below]. Nevertheless our qualitative conclusions should
not be altered by introducing evolution in the gluon distributions.

That way we see that the diffractive structure function in the large
$Q^2$ limit is linear in $x_{Bj} G$ with no $Q^2$ suppression. Due to
the multiple rescattering effects, the $F_2^D$ structure function
becomes a leading twist expression, i.e., it is not suppressed by
powers of $Q^2$. By resumming higher twist terms we obtained an answer
which is effectively leading twist. This result agrees with the
previous predictions of \cite{kope,bh}. Eq. (\ref{laq}) shows that the
diffractive structure function depends on energy in the same way as
the structure function $F_2$ at large $Q^2$. This result agrees with
the recent ZEUS data \cite{zeus}.

\section{Generalizing to Hadrons}

To generalize the considerations above to the case of a hadron, we
first recognize that the renormalization group treatment proposed in
\cite{kmmw,klmw,JKW} replaces the gluon distribution of a hadron by a
set of sources which include both valence quarks and sea gluons.  If
we go to small enough x, the gluon density per unit area is large
enough so that one can use a weak coupling analysis.  In this respect,
the situation is entirely parallel to that for a nucleus.

The essential difference is that the function which we use to average over
sources changes.  If we use a Gaussian distribution of sources,
\begin{eqnarray}
  \int [d\rho] \exp \left( - \int d^2 {\bf x} \frac{\rho^2 ({\bf
        x})}{2\Lambda (y)} \right)
\end{eqnarray}
then it is an easy exercise to show that the result of the last
section are reproduced.  To do this we use the fact that the factor
$\Lambda(y)$ is the charge squared per unit area [as defined in
\cite{MV1}] with $y \sim \ln x_+$ the space--time rapidity, and the
DGLAP equations \cite{dglap}. As was discussed in some detail by
McLerran and Venugopalan in \cite{MV1} the propagator of a single
quark through the nucleus can be represented as a path ordered
exponential (Wilson line)
\begin{equation}
P \exp \left( i \int_{x_+}^\infty d x'_+ A_- (x'_+) \right),
\end{equation}
where the field $A_-$ is the gluon field taken in covariant gauge.
Direct calculation of the path ordered integral done in \cite{MV1}
yields for the propagator of the quark--antiquark pair through the
hadron [see formula (119) in \cite{MV1}]
\begin{equation}\label{propg}
  \gamma ({\bf x}) = \exp \left[ \alpha^2 \pi C_F {\bf x}^2 \ln ({\bf
      x}^2 \Lambda_{QCD}^2) \, A \, \Lambda (y) \right],
\end{equation}
where the no interaction contribution was included in the propagator
in Eq. (\ref{propg}). $A$ is the atomic number of the nucleus. Using
the fact that
\begin{equation}
  A \Lambda (y) = \frac{\sqrt{R^2 - b^2} \rho}{C_F} \int_{x_{Bj}}^1 d
  x' \, G (x', 1/{\bf x}^2 )
\end{equation}
with $\rho = \frac{A}{(4/3) \pi R^3}$ the nucleon number density in
the nucleus we obtain
\begin{equation}\label{prp}
  \gamma ({\bf x}) = \exp \left( \alpha^2 \pi \sqrt{R^2 - b^2} {\bf
      x}^2 \ln ({\bf x}^2 \Lambda_{QCD}^2) \, \rho \int_{x_{Bj}}^1 d
    x' \, G (x', 1/{\bf x}^2 ) \right),
\end{equation}
To establish the correspondence between the result of Eq. (\ref{prp})
and the $q \overline{q}$ propagators from Sect. III one should make
use of the double logarithmic limit of the DGLAP equation
\begin{equation}\label{dglap}
  \frac{\partial}{\partial \ln (Q^2 / \Lambda_{QCD}^2)} x_{Bj} G
  (x_{Bj},Q^2 ) = \frac{\alpha N_c}{\pi} \int_{x_{Bj}}^1 d x' \, G
  (x', Q^2 ).
\end{equation}
In the quasi--classical calculation of the previous section the gluon
distribution function were taken only at the two gluon level.  Noting
that in that two gluon exchange approximation the gluon distribution
is given by
\begin{equation}\label{lo}
  x_{Bj} G (x_{Bj},Q^2 ) = \frac{\alpha C_F}{\pi} \ln
  \frac{Q^2}{\Lambda^2_{QCD}}
\end{equation}
one can easily see that Eq. (\ref{prp}) reduces to 
\begin{equation}
  \gamma ({\bf x}) = \exp \left( - \frac{\alpha \pi^2 \sqrt{R^2 -
        b^2}}{N_c} {\bf x}^2 \rho x_{Bj} G (x_{Bj}, 1/{\bf x}^2 )
  \right),
\end{equation}
which matches the exponents employed in derivation of Eqs.
(\ref{diffxs}) and (\ref{diffsf}) if one substitutes Eq. (\ref{vg})
into them. That way we showed that the result derived for the
structure function of a nucleus [Eq. (\ref{diffsf})] is, probably,
also valid for a hadron.

There are two assumptions here. First is that one can use the DGLAP
equations and the second is that the distribution of sources is local
and Gaussian.  These are true when one is at high enough momentum
transfer so that the density of gluon $dN/d^2p_Tdy$ per unit area is
small.  This distribution at large $p_T$ goes as $1/p_T^2$.  As the
transverse momentum is lowered therefore the density becomes large and
our assumptions break down.

The greatest complication in the small $p_T$ region ($p_T^2$ less than
or of the order of the color charge density of gluon per unit area) is
that the Gaussian distribution changes to a presumably non-Gaussian
and non-local (in transverse space) form.  In the exchanges of gluons
between the produced jet of quark-antiquark pairs, some transverse
momentum can be absorbed by the source.  For local Gaussian
distributions
\begin{eqnarray}
  \left< \rho({\bf x}_1) \ldots \rho({\bf x}_N)\right>
\end{eqnarray}
factorizes into a product of terms such as
\begin{eqnarray}
  \left< \rho({\bf x}_1) \rho({\bf x}_2) \right> \, \sim \,
  \delta^{(2)} ({\bf x}_1 - {\bf x}_2 ).
\end{eqnarray}
The locality of this product and its translational invariance are what
guarantee that the source absorbs no momentum.

If there is non-locality in the Gaussian averaging, then one couples
together sources at different transverse coordinates.  There will be
overall no transverse momenta transferred to the sum of sources, but
there may be momenta transferred to the individual sources at their
respective transverse coordinates.  (We have a picture in mind where
we have coarse grained in transverse coordinate.  Each coordinate cell
contains a separate source.)  If this is the case, then the Fock space
wavefunction of the hadron has been modified somewhat from its
original form as in the amplitude one has redistributed the transverse
momentum among its various constituents.

The problem has therefore become immensely more complicated.  One
should note however that this can only affect the soft transverse
momentum part of the hadron wavefunction, so that perhaps this does
not spoil the wavefunction so much.  It may also turn out that the
non-localities in the renormalization group improved distribution
function for sources is small.

As a practical matter, the transverse momentum scale of interest for
the non-linearities is presumably the virtuality of the photon $Q^2$.
If one is at large $Q^2$, then non-linearities and non-locality is not
important.  In this case, the exponential factors in Eq.
(\ref{diffsf}) linearize, and one is back to the simple two gluon
exchange model for diffractive processes.  Such a model has been
derived in the literature \cite{g2} from various considerations
different than those above.

\section{Total Inclusive Cross Section}

In the calculation outlined above we have derived the expressions for
diffractive (elastic) quark--antiquark pair production and structure
function in DIS on a nucleus in the quasi--classical approximation
[Eqs. (\ref{diffxs}),(\ref{diffsf})]. Our results include all higher
twist effects at the classical level. As was noted above the classical
limit in our case is understood as modeling the interactions by no
more than two gluons per nucleon. Quantitatively that results in
$\alpha_s^2 A^{1/3}$ being the expansion parameter. That way our
calculation could be understood as resumming all powers of this
parameter.

We should note here that similar results were obtained by
Buchm\"{u}ller, McDermott and Hebecker in \cite{BDH}. However, our
expressions in Eqs. (\ref{diffxs}),(\ref{diffsf}) give a more explicit
power of the exponential in the propagator of the pair through the
nucleus.

We should also note that using the techniques outlined above and used
previously in \cite{mM} one can also calculate the total (inelastic)
$q \overline{q}$ production cross--section. The $q \overline{q}$
propagator calculation is a little more difficult. In the framework of
the quasi--classical approach one has to also include one gluon
exchanges, which leave the nucleon in the color octet state, therefore
breaking it apart.  Without giving any details we state the results
\cite{KKMV}. The quark--antiquark total production cross--section is
\begin{eqnarray*}
  \frac{d \sigma}{d^2 k \, dz} = \frac{1}{\pi (2 \pi)^3} \, \int d^2
  b \, d^2 x \, d^2 y \,e^{- i {\bf k} \cdot ({\bf x} - {\bf y})} \,
  \Phi ({\bf x},{\bf y}, z )
\end{eqnarray*}
\begin{eqnarray*}
  \times \left[ 1 - \exp \left( - \frac{{\bf x}^2 \tilde{v} ({\bf x})
        \sqrt{R^2 - b^2}}{2 \lambda} \right) - \exp \left( -
      \frac{{\bf y}^2 \tilde{v} ({\bf y}) \sqrt{R^2 - b^2}}{2 \lambda}
    \right) \right.
\end{eqnarray*}
\begin{eqnarray}\label{totxs}
  + \left. \exp \left( - \frac{({\bf x} - {\bf y})^2 \tilde{v} ({\bf
  x} - {\bf y}) \sqrt{R^2 - b^2}}{2 \lambda} \right)\right],
\end{eqnarray}
where the transverse momentum of one of the quarks is fixed to be
${\bf k}$ and the momentum of the other is integrated over. This
formula reduces to a similar formula derived in \cite{BDH} [Eq.  (32)]
only in the logarithmic limit, when we can neglect the ${\bf x}$
dependence in $\tilde{v} ({\bf x})$. The nuclear structure function in
this quasi--classical limit is
\begin{eqnarray*}
  F_2 = Q^2 \frac{2 N_c}{\pi (2 \pi)^3} \sum_f \, Z^2_f \, \int d^2 b
  \, d^2 x \, \int_0^1 dz \left\{ a^2 K_1^2 (x a) [z^2 + (1 - z)^2] +
    m_f^2 K_0^2 (x a) + 4 Q^2 z^2 (1 - z)^2 K_0^2 (x a) \right\}
\end{eqnarray*}
\begin{eqnarray}\label{nucsf}
  \times \left[ 1 - \exp \left( - \frac{{\bf x}^2 \tilde{v} ({\bf x})
        \sqrt{R^2 - b^2}}{2 \lambda} \right) \right] ,
\end{eqnarray}
which agrees with the previously derived expressions [see
\cite{MV1,Mueller} and references therein].  In the limit of large
$Q^2$ Eq. (\ref{nucsf}) provides us with the usual leading twist
result
\begin{equation}
  \frac{\partial F_2}{\partial \ln (Q^2/Q_0^2)} = \frac{\alpha_s}{3
    \pi} \sum_f \, Z^2_f \, x_{Bj} G (x_{Bj}, Q^2).
\end{equation}

That way we have established a procedure of inclusion of higher twist
terms in the nuclear or hadronic structure functions and cross
sections in the quasi--classical approximation. In both cases of
diffractive and total inclusive cross sections the calculation is
simple and follows the same prescription. One has to convolute the
wave functions of a virtual photon with the propagator of the
quark--antiquark pair through the nucleus. For the case of total
inclusive cross section one should first square the propagator and
then average it in the nuclear wave function. In the case of
diffractive process one has to first average the propagator in the
nuclear wave function and then square it.

The relationship between the expressions (\ref{diffsf}) and
(\ref{nucsf}) can be shown in a different way \cite{MS}.  Let us
rewrite the total and elastic cross sections for the deep inelastic
scattering on a hadron or a nucleus in terms of the $S$-matrix at a
given impact parameter of the collision, $S(b)$, as \cite{MS}
\begin{mathletters}
\begin{eqnarray}\label{stot}
\sigma_{tot} = 2 \int d^2 b \, \, [1 - S(b)] 
\end{eqnarray}
\begin{eqnarray}\label{sel}
\sigma_{el} =  \int d^2 b \, \, [1 - S(b)]^2 .
\end{eqnarray}
\end{mathletters}
Now one can see that the diffractive (quasi--elastic) cross sections
of a quark--antiquark pair interacting with the nucleus, which was
used in Eq. (\ref{diffsf}), corresponds to the formula (\ref{sel}),
whereas the total inclusive cross section used in formula
(\ref{nucsf}) corresponds to Eq. (\ref{stot}). Eqs.  (\ref{stot}) and
(\ref{sel}) show how easily the quasi--elastic and and total cross
sections could be obtained from one another.

\section*{Acknowledgments}

The authors would like to thank Genya Levin, Alfred Mueller and Raju
Venugopalan for many insightful suggestions and commentaries. We thank
the anonymous referee for a particularly perspicacious comment which
led us to understanding that $F_2^D$ is a leading twist effect. We are
grateful to Urs Wiedemann for pointing out a mistake in the earlier
version of the paper. This work is supported by DOE grant
DE-FG02-87ER40328.

\end{document}